# Signature of the Overhauser field on the coherent spin dynamics of donor-bound electron in a single CdTe quantum well

C. Testelin[1], F. Bernardot[1], G. Karczewski[2], and M. Chamarro[1]

*(1) Institut des NanoSciences de Paris - Universités Paris-VI et Paris-VII, CNRS UMR 7588, 4 place Jussieu, 75252 Paris cedex 05, France*
*(2) Institute of Physics, Polish Academy of Sciences, Al. Lotnikow 32/46, 02-668 Warsaw, Poland*

**Abstract:** We have studied the coherent spin dynamics in an oblique magnetic field of electrons localized on donors and placed in the middle of a single CdTe quantum well, by using a time-resolved optical technique: the photo-induced Faraday rotation. We showed that this dynamics is affected by a weak Overhauser field created *via* the hyperfine interaction of optically spin-polarized donor-bound electrons with the surrounding nuclear isotopes carrying non-zero spins. We have measured this nuclear field, which is on the order of a few mT and can reach a maximum experimental value of 9.4 mT. This value represents 13 % of the maximal nuclear polarization, and corresponds also to 13 % of maximal electronic polarization.

**Introduction**

The coherent dynamics of localized electron spins in semiconductor nanostructures is of great interest in the new fields of spintronics and quantum information [1-3]. Localization on the nanometer length scale strongly suppresses spin flip mechanisms active in bulk and 2D systems. Recent studies have shown that the coherence time of electron spins can be enhanced by at least two orders of magnitude when free electrons are localized on donors immerged inside a quantum well (QW) [4]. In the case of donors, at low temperature, the potential well which confines the electron is created by the coulombic interaction between the electron and the ionized donor atom. Electrons localized on donors are a reproducible and very homogeneous "model system" of n-doped quantum dots (QDs), for which the hyperfine interaction of the electron with the nuclear moments of the lattice atoms sets an ultimate limit for the spin relaxation time at zero magnetic field. The hyperfine interaction can also be at the origin of a polarization of the nuclear spins, which then build up an effective magnetic field, the Overhauser field. Such a dynamical nuclear polarization by electron spins has been largely observed in the past in bulk semiconductors [5-8] and QWs [9-12], and more recently in QDs [13-15] under continuous electron spin pumping. It induces singular effects on the behaviour of the electron spins, such as non-linear or bistable response to applied external fields [16-18]. Although the hyperfine interaction in III-V nanostructures has been more largely addressed, the hyperfine interaction in II-VI nanostructures deserves attention because it has been recently identified to be at the origin of a fast establishment of a dynamical polarization in II-VI QDs [15].

In this paper we report on the precise measurement of weak nuclear Overhauser fields obtained under pulsed excitation and *via* the hyperfine interaction of an electron on a donor placed in the centre of an 80Å QW with the surrounding active nucleus isotopes of CdTe. Here we use a pump-probe technique, the photo-induced Faraday rotation (PFR), to polarize the localized electronic spin by a resonant optical pumping of free excitons, and to monitor its dynamics. The appearance of a nuclear magnetic field, when a magnetic field is applied out of the direction of the exciting light, affects the coherent spin dynamics of the localized electrons. Indeed, in an oblique magnetic field the optically created electronic spin can be represented by two orthogonal components: a precessing component perpendicular to the applied magnetic field, and the other one parallel to the applied magnetic field and responsible for the appearance of the nuclear polarization. Then, the transverse component of the electronic spin precesses in a total magnetic field which is the result of the algebraic addition of both nuclear and external fields. The measurement of the Larmor frequency allows the determination of the induced nuclear field.

**Sample characterization**

The studied sample consists of a CdTe/CdMgTe heterostructure grown by molecular-beam epitaxy on a (100)-oriented GaAs substrate and containing an 80 Å QW. A donor layer of iodine atoms was placed at the middle of the QW. The donor concentration is approximately $10^{11}$ cm$^2$. In order to perform transmission and PFR measurements we have chemically suppressed the GaAs substrate. The inset of Fig. 1 shows the low-temperature photoluminescence (PL) spectrum of the studied sample, obtained with a 5 mW cw excitation at 633 nm. It is dominated by a peak at 1.617 eV, which is attributed to the recombination of the three-particle complex $D^0X$ formed by a donor-bound exciton. The free exciton recombination is also observed in the PL spectrum as a shoulder at 1.621 eV.

A train of 2-ps pulses from a mode-locked Ti:sapphire laser with a 76-MHz repetition rate is split into the pump and probe beams. In this degenerate configuration, the pump average intensity was in the order of 1 W/cm$^2$, and the probe average intensity was ten times less. The pump beam is circurlarly polarized, and the probe beam is linearly polarized. After transmission through the sample, we measure the rotation angle of the probe beam polarization by using an optical bridge. The pump and probe beams are modulated with an optical chopper at 1 KHz and 1.3 KHz, respectively, and the signal of the probe beam is recorded with a lock-in amplifier. More experimental details are given in previous publications [19]. Figure 1 shows the PFR signal at zero magnetic field as a function of the pump-probe delay, for two different excitations: a) a resonant excitation at the free exciton energy, and b) a resonant excitation of the $D^0X$ complex. We underline that both curves show a non-zero signal at negative pump-probe delay times, which indicates that the spin polarization is not fully relaxed within the 13-ns repetition period of the laser pulses. As the lifetimes of the photo-generated species, *i.e*, free excitons [20] and donor-bound excitons [21], are less than several hundreds of picoseconds, we can conclude that the electrons bound to donors acquire a net spin polarization, and that the probe beam is sensitive to the polarization of localized electrons. In similar samples the spin relaxation time of the donor-bound electron has been determined to be equal to 20 ns [4].

Both PFR curves of Fig. 1 are obtained for a circularly polarized pump pulse, which is absorbed at t = 0 and generates a transient population of free excitons or donor-bound excitons with a spin polarization parallel (resp.: antiparallel) to the incident beam for a σ+ (resp.: σ–) polarization. After capture, spin-relaxation and recombination processes, an unbalanced population of spin-polarized donor-bound electrons is obtained. The short-time components of the PFR signals should then be related to the free and donor-bound exciton spin dynamics [22], but they are not discussed in the following. Concerning the mechanism of the spin polarization of the resident electrons, it depends on the energy of the exciting beam. This spin polarization is built *via* the creation of donor-bound excitons, resonantly or indirectly by the capture by donors of resonantly excited free excitons. This capture process is very efficient, as demonstrated by the free-exciton recombination signal in the PL spectrum, which is much smaller than the $D^0X$ peak. In both cases, the essential condition for the realization of a spin polarization of the localized electrons is that the spin relaxation time of the photo-created hole bound to $D^0X$ be in the same order of magnitude or smaller than the recombination time of $D^0X$ [4, 19].

The time evolution of the net spin polarization is monitored through the rotation of the polarization plane of the transmitted probe pulse, which is almost collinear with the pump. In the case of the spin of localized electrons, when the pump and probe are tuned to the $D^0X$ formation energy, the rotation is clearly proportional to the population difference of electron spin states *via* optical transitions involved in absorption and transmission of the pump and probe beams. However, when the pump and probe beams are tuned to the energy of the free exciton formation, the probe is sensitive to the population difference of donor-bound electron spins *via* the exchange interaction of electrons contained in excitons created by the probe beam with the electrons localized on donors. This exchange interaction is inversely proportional to the detuning between the probe beam and the $D^0X$ transition [23]. In the following, we have chosen to work with the pump and probe energy tuned to the free exciton transition, in order to get the largest signal from the spin-polarized bound electrons. This situation arises from an oscillator strength larger for free excitons than for bound excitons [24], and from a very efficient capture of free excitons by donors.

When a transverse magnetic field is applied, the electron spin precesses around the field and the PFR signal shows damped oscillations (see Fig. 2a). In Fig. 2b) a scheme of our sample holder is given: two permanent magnets (grey rectangles) are placed on both sides of the sample to create an in-plane-of-QW external field of 0.29 T. The PFR signal is fit to a damped cosine

$$S_z \propto n_\uparrow - n_\downarrow \propto e^{-t/T_2^*} \cos(\Omega t + \varphi) , \quad (1)$$

where Ω is the Larmor frequency, fixed by the magnetic field B and the Landé factor $g_e^\perp$: $\Omega = g_e^\perp \mu_B B$. This fit gives $\varphi = 0$, $|g_e^\perp| = 1.3$ and the dephasing time $T_2^* = 5$ ns. The value of $|g_e^\perp|$ is in good agreement with other authors [25, 26].

**Nuclear polarization: results and discussion**

As discussed above, a pulse of pump beam creates a spin-polarized bound electron population, whose relaxation time is comparable or slightly larger than the repetition period of the laser pulses. Due to the contact hyperfine interaction, a nuclear spin polarization builds up by integration over many laser pulses of the mutual spin flip-flops of the bound electrons and the surrounding lattice nuclei. The nuclear magnetic field modifies the coherent electron spin dynamics, and the PFR is used here as a very sensitive technique to measure this Overhauseur field associated to polarized nuclear spins. By changing the beam incidence to an angle $\theta \neq \pi/2$, the nuclear field is driven out of the direction of the pump beam, and two components of the net spin polarization of the electrons appear: one perpendicular and one parallel to the applied magnetic field (see Fig. 2c). The parallel component does not precess and is responsible for the photo-induced nuclear field, meanwhile the perpendicular component precesses in a total magnetic field made of the external and nuclear fields. Thus, in the chosen experimental configuration, the signature of the Overhauser magnetic field appears in the Larmor frequency of the electron spin precession. Depending on the σ+ or σ– polarization of the pump beam, one Larmor precession is due to a total field B + B$_N$, and the other one to a field B – B$_N$, where B$_N$ is the Overhauser field. Figure 3 clearly shows that for an angle of incidence θ = 30°, the Larmor frequencies are different for a σ+ or σ– polarization of the pump beam. For small pump-probe delays, the oscillations are in antiphase, as expected, but they appear to be at different frequencies near delay 1.2 ns; the frequency difference is then measured with the oscillations observable at negative delay times, corresponding to delay times around the 13-ns period of the laser. The difference of Larmor frequencies is related to B$_N$ by the following expression:

$$|B_N| = \frac{\hbar}{2|g_e^\perp|\mu_B} |\Omega_+ - \Omega_-|. \quad (2)$$

Figure 4 shows the experimental values of the difference of Larmor frequencies as a function of the angle of incidence θ. From Fig. 2c we can write that the parallel component of the electronic spin is:

$$S_\parallel = S_z \sin\alpha = S_z \sin\theta/n, \quad (3)$$

where α is the internal refraction angle and $n$ is the refractive index of the CdTe QW ($n = 3.3$); $z$ is the axis along the pump beam inside the sample. The related nuclear field then writes

$$B_N = \frac{B_N^0}{n} \sin\theta, \quad (4)$$

where $B_N^0$ is the maximal Overhauseur field that would be obtained if the pump beam were in the plane of the QW. Figure 4 also shows experimental values of $B_N$ obtained for angles of incidence between –30° and 40°. As we will discuss later, the experimental $\mathbf{B}_N$ is parallel to the applied magnetic field for a left-circularly polarized pump and a positive angle of incidence, as shown in the inset of Fig. 4. The solid line is a fit of the experimental data to Equation (4), which gives a maximal Overhauseur field $B_N^0 = 9.4$ mT. It is worth noting that this nuclear polarization is at least two orders of magnitude weaker than the value obtained in III-V semiconductors [8,12], and is in good agreement with measurements in bulk p-doped CdTe under cw excitation [7].

The Overhauser field $B_N$ is also defined by intrinsic parameters characterizing the material and the hyperfine interaction in the material: when all magnetic Cd and Te nuclei are oriented, the nuclear field writes [8]:

$$B_N^{max} = \frac{\sum_i I^i A^i p^i}{g_e^\perp \mu_B}, \quad (5)$$

where $i$ = Cd or Te, $p^i$ is their abondances ($P_{Cd} = 25\%$ and $P_{Te} = 8\%$) and $I^{Cd} = I^{Te} = 1/2$ is the non-zero nuclear spin. $A^i$ is the hyperfine constant, which can be calculated from the following expression [8]:

$$A^i = \frac{16\pi \mu_B \mu_I^i |u_i(0)|^2}{3 I^i}, \quad (6)$$

where $u_i(0)$ is the Bloch amplitude at the site of the nucleus and $\mu_I^i$ is the magnetic moment of a given nucleus ($\mu_I^{Cd} = -0.6077 \mu_B$, $\mu_I^{Te} = -0.8703 \mu_B$). By taking the experimental value of $|u(0)|^2$ found by Nakamura *et al.* [7], we calculate $A_{Cd} = -31$ μeV et $A_{Te} = -45$ μeV. Using Equation (5) we then obtain an estimated $B_N^{max} \approx 75$ mT. By comparison of this value with the experimentally determined $B_N^0 = 9.4$ mT, we conclude that a degree of nuclear orientation of 13 % would be achieved in a pulse regime for a pump average intensity of 1 W/cm² and an angle of incidence of 90°. To make a comparison with nuclear polarizations obtained under cw excitation, we have to take into account that the electronic polarization relaxes with a characteristic time of $\tau_e = 20$ ns [4] and that the pulse period is $T_L \approx 13$ ns; assuming an exponential decay for the electronic polarization, we obtain that a pulsed excitation reduces the nuclear polarization by a factor

$$\int_0^{T_L} e^{-\frac{t}{\tau_e}} dt / T_L = \frac{\tau_e}{T_L}[1-\exp(-T_L/\tau_e)] \approx 0.74, \quad (7)$$

as compared to a cw optical excitation. Then the maximum nuclear polarization that takes into account this factor is 13/0.74 = 17.5 %, which is higher than the 8 % reported in bulk p-doped CdTe [7]. Notably, these nuclear polarization rates remain lower than those found in InAs QDs and in charge-tunable GaAs interface QDs, where the reached nuclear polarizations are 40 % and 60 %, respectively, under cw optical excitation.

The experimental conditions insure the high temperature of the nuclear spins, *i.e.*, $\langle I \rangle / I \ll 1$, and then we can write [8]:

$$\langle I \rangle = \frac{I(I+1)}{S(S+1)} \langle S \rangle. \quad (8)$$

In our case I = 1/2 and S = 1/2. Equation (8) then becomes $\langle I \rangle = \langle S \rangle$, and allows to know the average electronic polarization when the nuclear polarization is known. In our experimental configuration, the maximal electronic polarization, in the case of a maximal creation of bound excitons and of a cw experiment, is given by the following expression:

$$\frac{\langle S \rangle}{S} = \frac{\gamma_h}{2(\gamma_h + \gamma_R)}, \quad (9)$$

where $\gamma_h$ and $\gamma_R$ are the spin relaxation rate for the photo-generated hole and the radiative rate of the $D^0X$ complex, respectively. For our sample $\gamma_h \approx \gamma_R$ and the maximal expected value of the electronic polarization is 25 %, which is slightly larger than the estimated 17.5 % for an non-optimized pump average intensity of 1 W/cm$^2$.

As the Landé factor $g_e^\perp$ is negative [25] and the hyperfine constants are negative because the $\mu_I$ are negative, the sign of $B_N$ is fixed by the sign of $\langle I \rangle$ which is set by the sign of $\langle S \rangle$. For a left-circularly polarized pump and a positive angle of incidence, $\mathbf{B_N}$ is parallel to the applied magnetic field, as shown in the inset of Fig. 4. A reversal of angle of incidence or polarization of the pump beam reverses $\langle S \rangle$ and therefore also $\langle I \rangle$, giving a nuclear field $\mathbf{B_N}$ antiparallel to the applied field.

**Conclusion**

In conclusion, we have demonstrated that the coherent spin dynamics of donor-bound electrons inside a single CdTe QW, is affected by the building of a very weak nuclear field due to the hyperfine interaction. We have also shown that the PFR technique is very sensitive and allows the determination of a nuclear field on the order of few mT, at least two orders of magnitude weaker than the value obtained by other authors in III-V semiconductors, and more than twice the already reported value in bulk CdTe [7]. Moreover, due to the fact that our experimental conditions insure the high temperature of nuclear spins, we are able to evaluate the electronic polarization from the nuclear polarization. We show that our estimation, in the order of 13 %, is in agreement with simple theoretical arguments.

**Acknowledgements**

The authors acknowledge financial support of the Île-de France Regional Council through the "project SESAME 2003" n°E. 1751.

**Figure captions**

**Figure 1** Inset: low-temperature PL spectrum obtained with a 5-mW cw excitation at 633 nm. Decay of the PFR signal obtained for two different excitation energies: a) The pump and probe energy is tuned to the energy of the free exciton recombination (1.621 eV); b) The pump and probe energy is tuned to the energy of the $D^0X$ recombination (1.617 eV).

**Figure 2**

a) Low-temperature difference of the PFR signals obtained in a transverse magnetic field 0.29 T for the two helicities σ+ and σ– of the pump beam. The energy of the pump and probe beams is tuned to the free exciton recombination energy.
b) Scheme of the sample holder. Two permanent magnets (rectangles in grey) create an in-plane magnetic field of 0.29 T at the place of the sample (square).
c) Experimental configuration used to achieve an oblique magnetic field with an incidence angle θ.

**Figure 3** PFR signals obtained at θ = 30°, B = 0.29 T, for a pump and probe energy tuned to the free exciton recombination. The dashed and solid curves correspond to the two different helicities of the circularly polarized pump beam. The thin dashed horizontal lines represent the zero-signal levels. Inset: magnification of both curves near the 13-ns delay time, *i.e.*, at negative delays; the arrows show maxima of the dashed curve that, in absence of the induced nuclear field, should correspond to minima of the solid curve.

**Figure 4** Difference of the Larmor frequencies for a pump beam circularly polarized σ– and σ+, and corresponding nuclear magnetic field, as a function of the angle of incidence θ. The line is a fit to the expected function given by Equation (4), see text. Inset: scheme in which the direction of the optically injected electronic spin and the induced nuclear field $B_N$ are given for a pump beam left-circularly polarized.

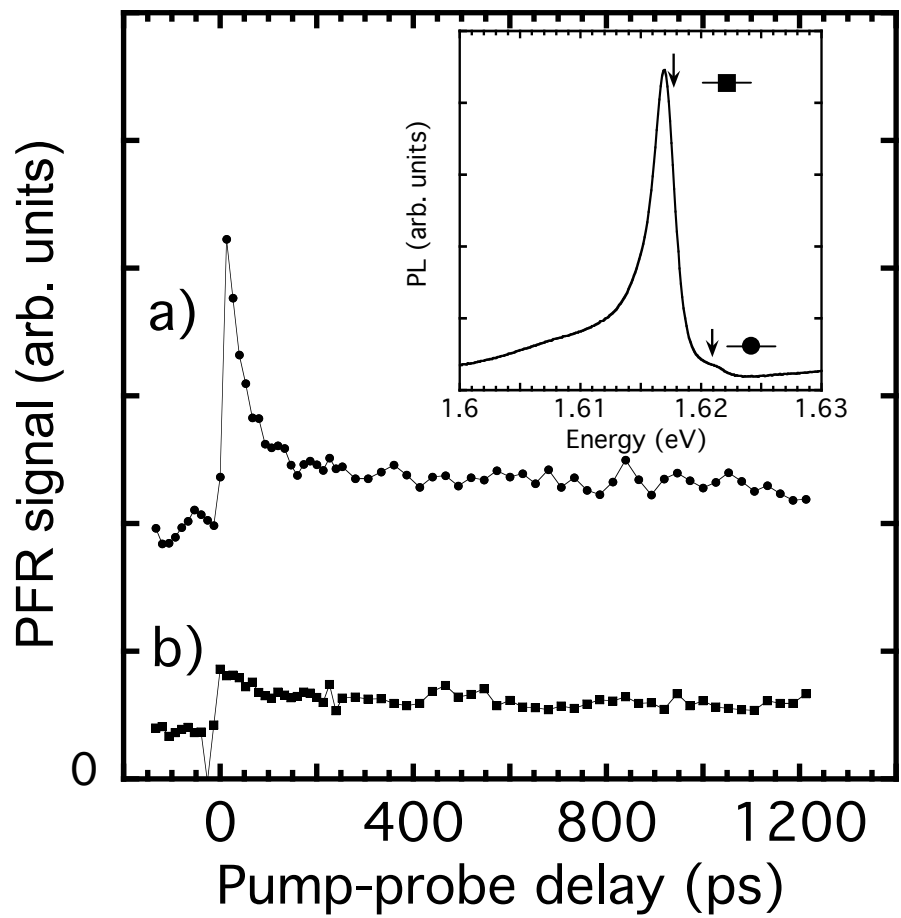

Figure 1

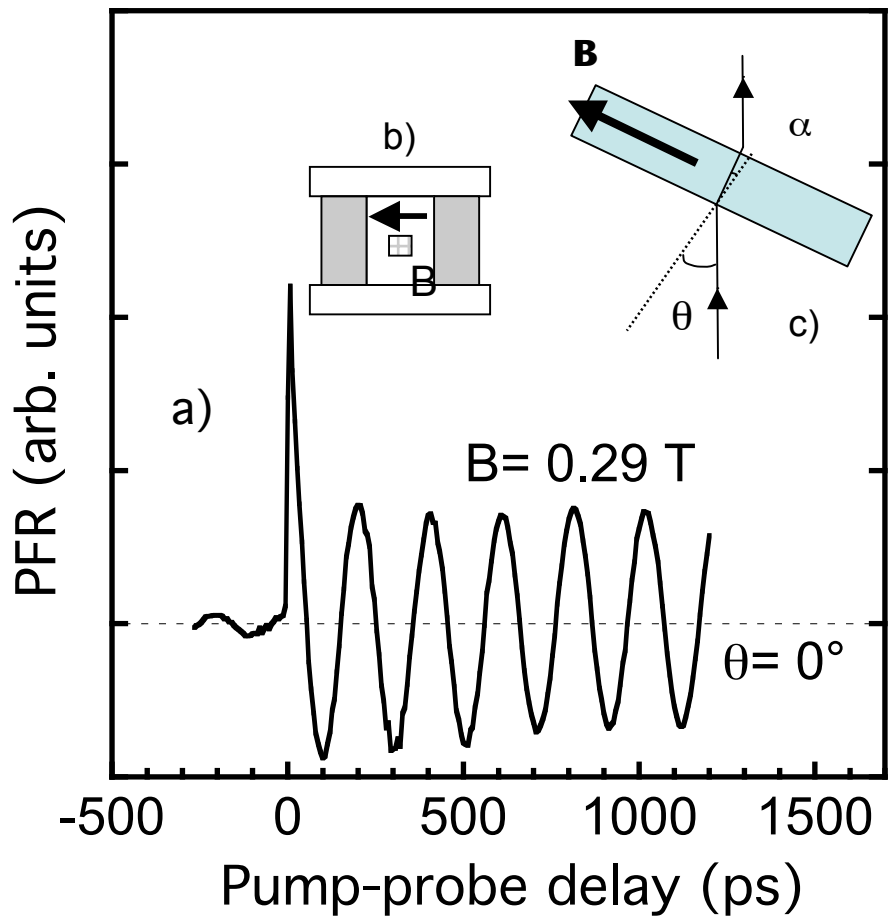

Figure 2

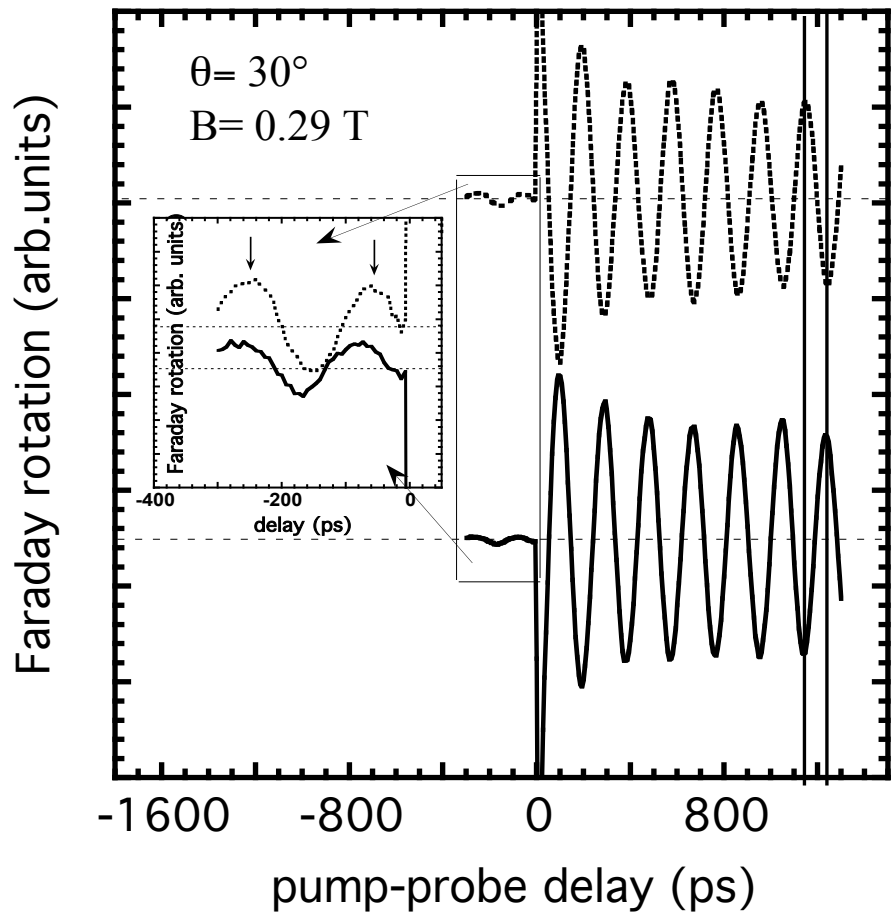

Figure 3

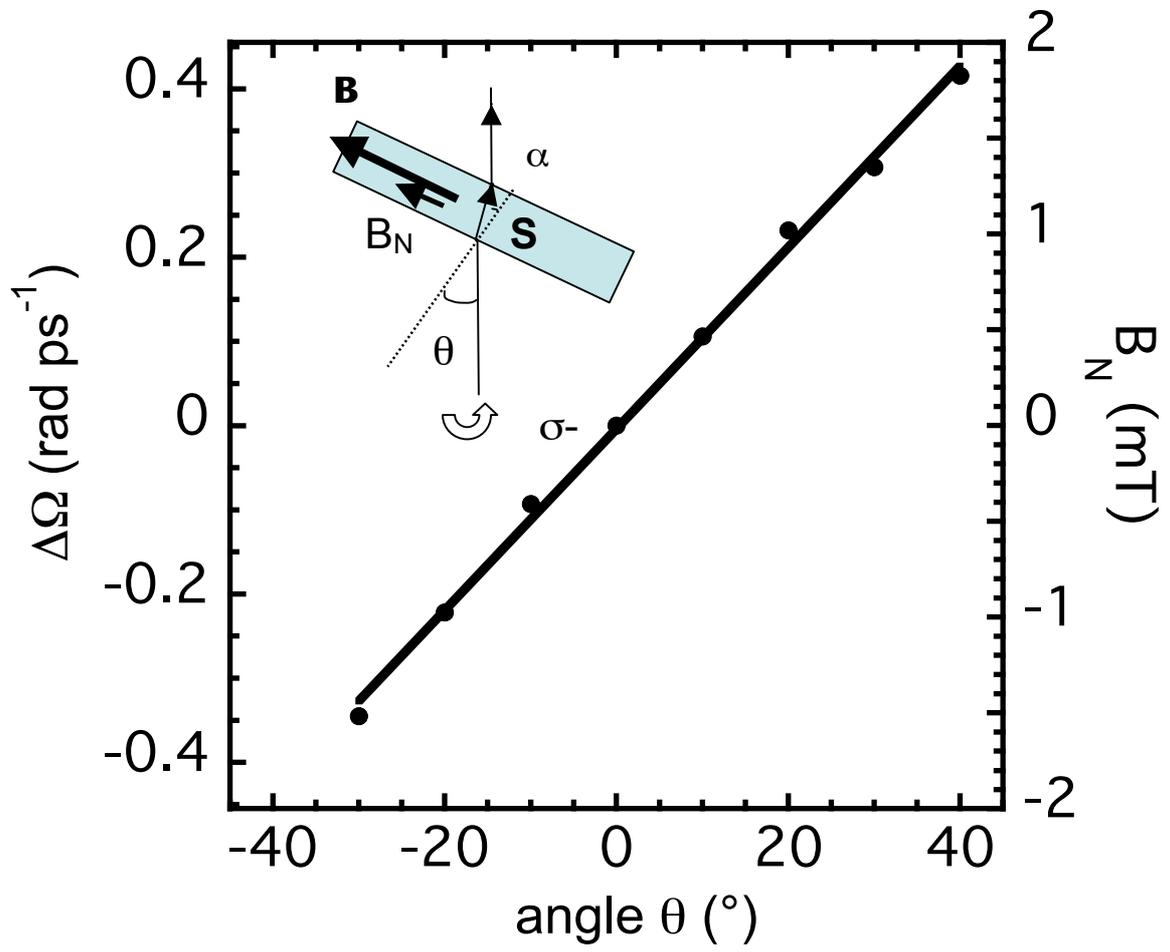

Figure 4